\documentclass{emulateapj}
\usepackage{graphicx}
\usepackage{amsmath}
\usepackage{natbib}
\usepackage{lscape}

\shorttitle{SNe Ib/c 2004ao, 2004gk and 2006gi}
\shortauthors{Elmhamdi et al.}
\slugcomment{Accepted for publication in \emph{The Astrophysical Journal}}

\begin{document}
\title{Photometric Evolution of SNe Ib/c 2004ao, 2004gk and 2006gi}

\author{Abouazza Elmhamdi\altaffilmark{1}, 
  Dmitry Tsvetkov\altaffilmark{2},
  I. John Danziger\altaffilmark{3,4},
  and Ayman Kordi\altaffilmark{1} }

\altaffiltext{1}{Department of Physics and Astronomy, College of Science, King Saud University. PO Box 2455, Riyadh 11451, Saudi Arabia.}

\altaffiltext{2}{Sternberg Astronomical Institute, University Ave. 13, 119992 Moscow, Russia.}

\altaffiltext{3}{INAF, Osservatorio Astronomico di Trieste, Via G.B. Tiepolo 11, 34131 Trieste, Italy.}

\altaffiltext{4}{Department of Astronomy, University of Trieste, Via G.B. Tiepolo 11, 34131 Trieste, Italy.}

\begin{abstract}
 
 Photometric observations of three core collapse supernovae (SNe 2004ao, 
 2004gk and 2006gi), covering about 200 days of evolution are presented and 
 analyzed. The photometric behaviour of the three objects is consistent with 
 their membership of the envelope-stripped type Ib/c class. 
 Pseudo-bolometric light curves are constructed.
 The corresponding measured 
 $e$-folding times are found to be faster compared to the $^{56}$Co decay 
 (i.e. 111.3 d), suggesting that a proportion of $\gamma$-rays increasing 
 with time have escaped without thermalization, owing to the low mass nature 
 of the ejecta. 
 SN 2006gi has almost identical post maximum decline phase luminosities 
 as SN 1999ex, and found to be 
 similar to both SNe 1999dn and 1999ex in terms of the 
 quasi-bolometric shape,  
 placing it among the fast decliner Ib objects. 
 SN 2004ao appears to fit within the slow decliner Ib SNe. 
 SNe 2004ao and 2004gk display almost identical
 luminosities in the [50-100] days time interval, similar to SN 1993J. 
 A preliminary simplified $\gamma -$ray deposition model is described and 
 applied to the computed pseudo-bolometric light curves, allowing one to find
 a range in the ejecta and $^{56}$Ni masses. 
 The optical and quasi-bolometric light curves, and the $B-V$ colour 
 evolution of SN 2004gk are found to show a sudden
  drop after day 150. Correlating this fact to dust formation is premature
  and requires further observational evidence. 
\\
\end{abstract}


\section{INTRODUCTION}

Stripped-envelope SNe (SESNe)commonly display variety in their
photometric evolution (Clocchiatti and Wheeler. 1997; Richardson, Branch
 and Baron. 2006). The heterogeneity in this class of events is also imprinted
in their spectra (Elmhamdi et al. 2006; Taubenberger et al. 2009;
 Maurer et al. 2010). These facts together with the appearance 
 of highly energetic objects with extremely broad spectral features, 
 labeled `Hypernovae', and the established 
 association with the long-timescale Gamma-Ray Bursts have stirred 
 much interest in the scientific community during the past decade.

 As far as the photometry is concerned, extensive studies in the literature 
 have argued that the shape of the type Ib/c light curves and the way
 they decline from the peak are related
 to the the progenitor star properties, in particular the ejected mass, 
 explosion energy, the progenitor radius and the  
 degree of mixing (Arnett 1982; Ensman \& Woosley 1998). 
 The smaller the ejected mass to explosion energy ratio $M_{ej}/E_k$ 
 and the closer the $^{56}$Ni to the 
 surface, the faster the light curve declines (Shigeyama et al. 1990). One 
 scenario for type Ic SNe indicates in fact that they originate from slightly 
 smaller masses at the time of explosion compared to type Ib objects 
 (Yamaoka $\&$ Nomoto 1991). For type Ic SNe, two subclasses
 having different light curve behaviour are observed, namely  fast and
 slow decliners (Clocchiatti and Wheeler 1997). The events like 1990B, 1992ar
 and the bright 1998bw 
 belong to the slow declining subgroup. SN IIb 1993J
 has similar behaviour at the post-maximum phase as the slow Ic class. 
 SNe 1983V, 1983I, 1987M and the well observed 1994I are fast declining 
 objects. Two points of interest should be mentioned in this respect.
 First, the brightest Ic objects display broader peak widths and smaller
 peak-to-tail contrast values with a greater delay in the occurrence of
 maximum light. Second, some of type Ic, such as SNe 2002ap and 2004aw, appear 
 to belong to an intermediate class between the fast Ic and the slow Ic 
 events. Whether a continuity exists in the Ic SNe variety from the 
 fast-less bright to the slow-bright objects is questionable.
 Worthy of note here is that also for type Ib light curves, the observed
  homogeneity, i.e being all slow decliners, is broken by cases such as 
 SN Ib 1991D (Benetti et al. 2002),
 SN 1990I (Elmhamdi et al. 2004) and SN 1999dn (Benetti et al. 2010) 
 showing a fast light curve decline from maximum light. 
 The diversity within SESNe subclass hence increases with our ability
 to detect and study them. Although still few, sample comparative
 studies are a potential tool for a major physical understanding of SESNe.
%%%%%%%%%%%%%%%%%%%%%%%%%%%%%%%%%%%%%%
\begin{figure*}
\centering
 \includegraphics*[width=5.5cm,height=5.5cm]{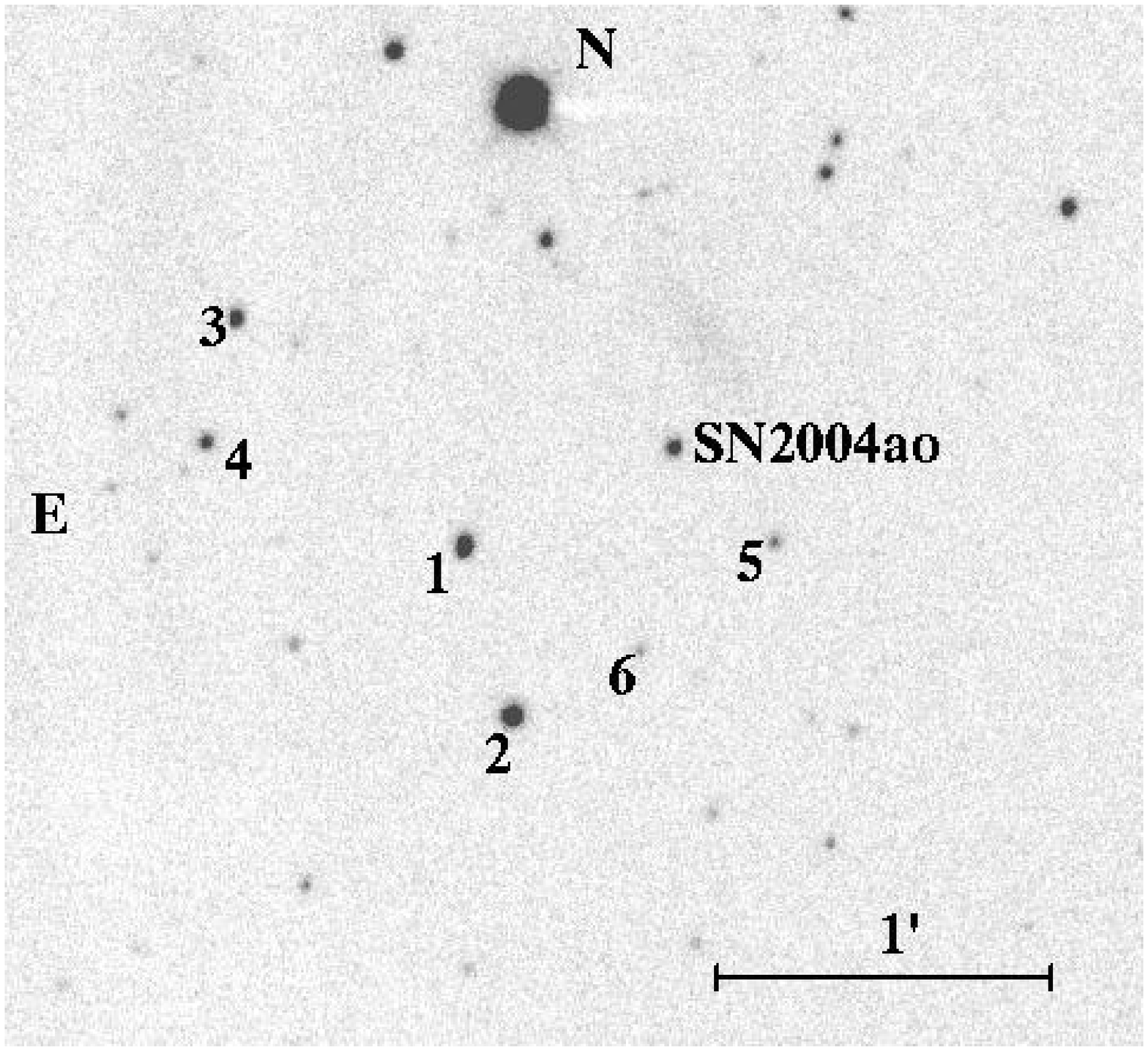}$~~$\includegraphics*[width=5.5cm,height=5.5cm]{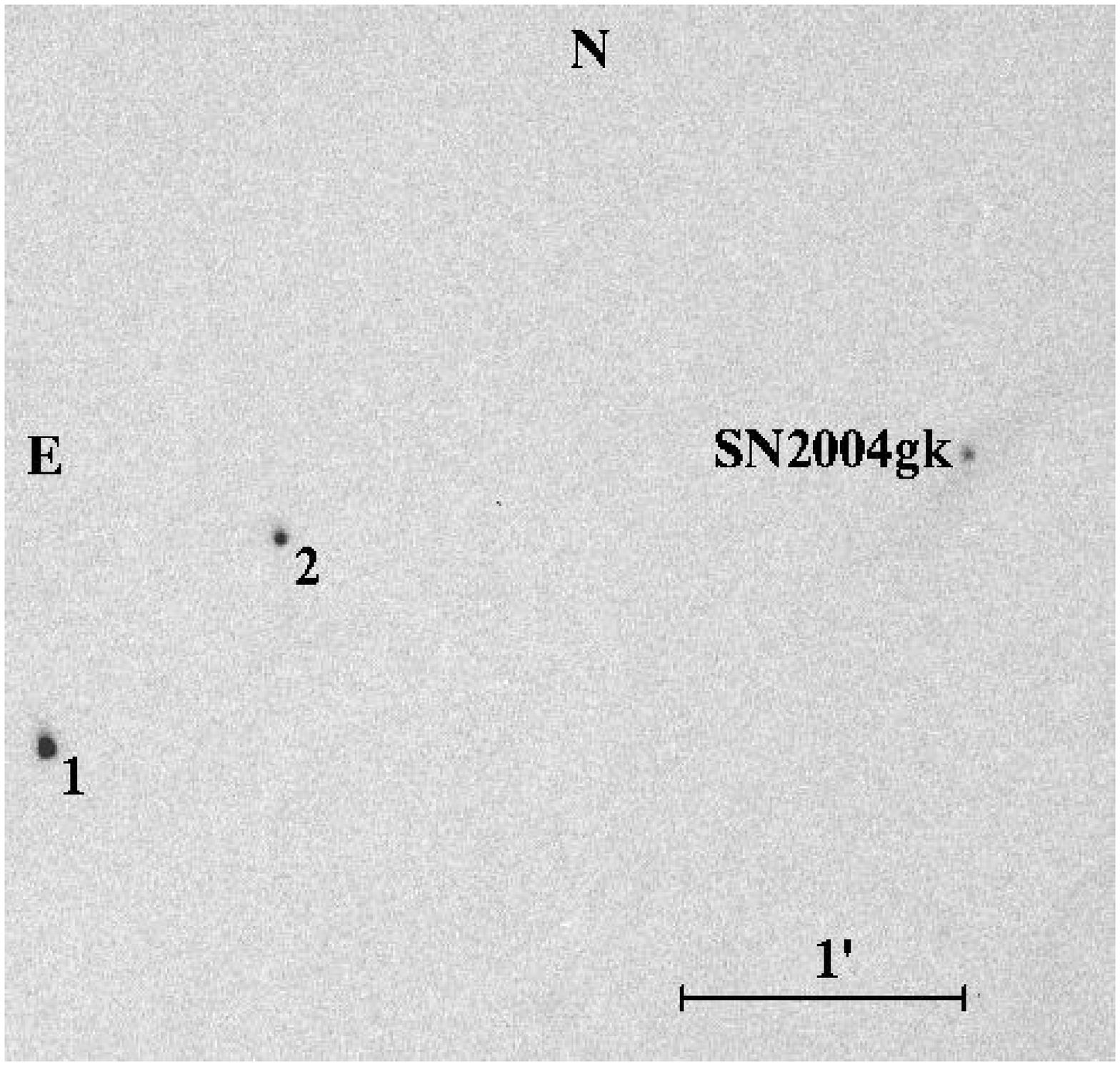}$~~$\includegraphics*[width=5.5cm,height=5.5cm]{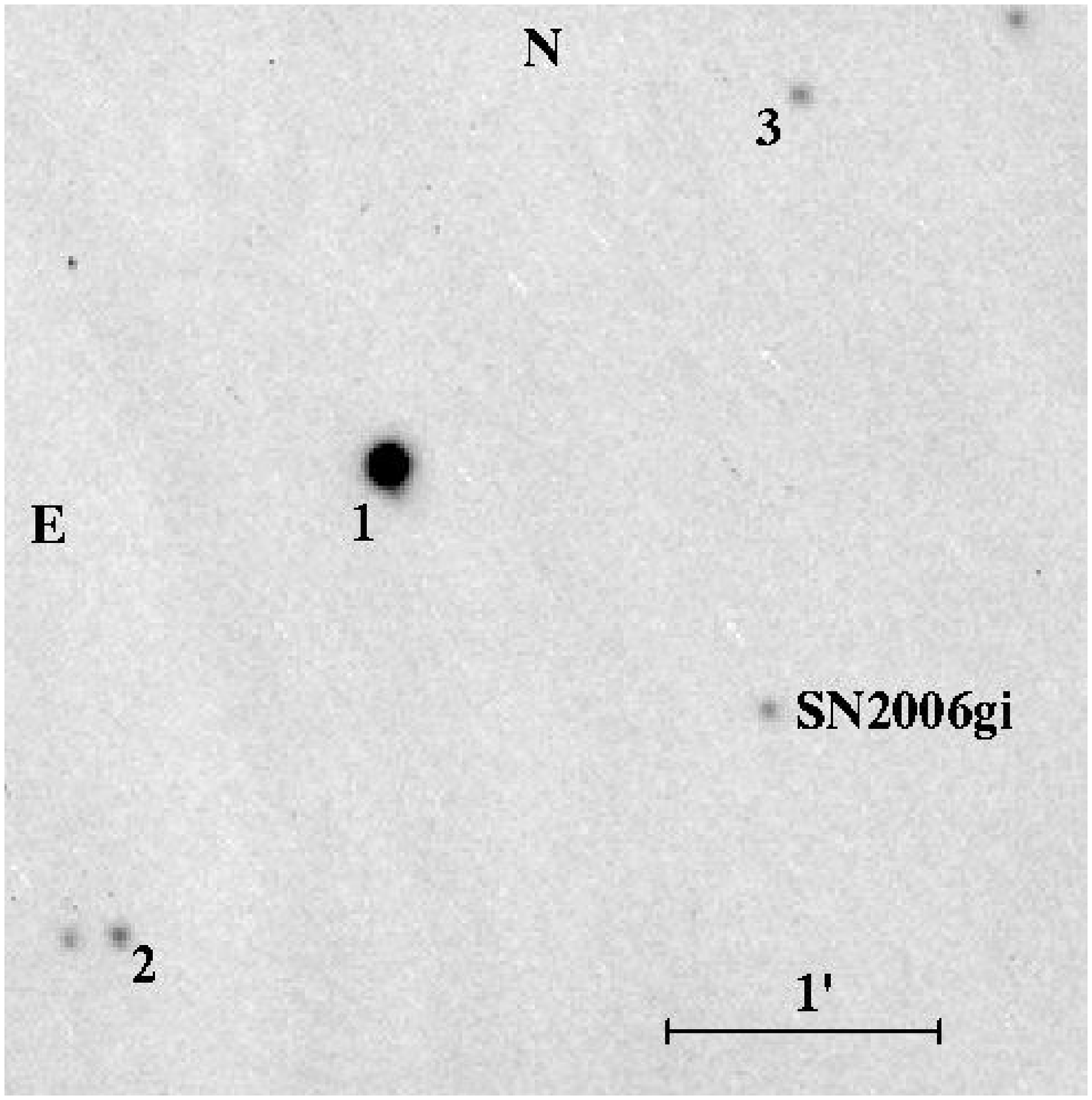}
 \figcaption{The $V$-band images showing the field of SNe 2004ao (left), 
 2004gk(middle) and 2006gi (right) and their local standards.}
 \end{figure*}
%%%%%%%%%%%%%%%%%%%%%%%%%%%%%%%%%%%%%
 
 The main goal of the present work is to contribute
 to enriching the observational set of SESNe data. We present
 and analyze original photometry observations of three events namely, 
 SNe 2004ao(Ib), 2004gk(Ic) and 2006gi(Ib). Until now only few data for these
 events, mainly spectra, have been published. 
 Some late phase spectra of 
 SNe 2004ao and 2004gk have been presented by Modjaz et al.
 (2008). By means of analyzing properties of the nebular lines the authors 
 report a double-peaked [O I] 6300,6364 \AA$~$ profile in SN 2004ao. 
 The SN appears to be peculiar in exhibiting a decreasing
 intensity ratio, [O I] 6300\AA$~$ to [O I] 6364 \AA, contrary to what is
 expected for the optically thin limit. 
 SN 2004gk is found to show 
 single-peaked [O I] 6300,6364 \AA$~$profile. The authors argued that
 asphericities are more likely present in a wide variety of SESNe. For SN 
 2006gi, Taubenberger et al. 2009 provide a nebular spectrum of the
 object among a sample of SESNe late spectra. The spectrum shows emission 
 lines typical for a type Ib object, with single-peaked 
 [O I] 6300,6364 \AA$~$profile.
   
 The present paper is organized as follows.
 In Section 2, we describe the observations and data reduction
 techniques. Broad-band photometry 
 data are presented and analyzed in Section 3. In Section 4 
 we discuss the light curves and colour evolution.
 The pseudo-bolometric light curves are constructed in Section 5, 
 and then compared with the computed bolometric light curves
 of other SESNe events. By means of a simplified $\gamma -$ray deposition 
 model we recover the ejecta and $^{56}$Ni masses after discussing restrictions
 on the kinetic energies. We conclude with a summary in Section 6.  

%%%%%%%%%%%%%%%%%%%%%%%%%%%%%%%%%%%%%
\section{OBSERVATIONS AND DATA REDUCTION}
%%%%%%%%%%%%%%%%%%%%%%%%%%%%%%%%%%%%%%

%%%%%%%%%%%%%%%%%%%%%%%%%%%%%%%%%%%%%
\begin{deluxetable}{lllllllll}
\tablewidth{0pt}
\tabletypesize{\scriptsize}
\tablecolumns{9}
\tablecaption{Magnitudes of comparison stars}
\tablehead{\colhead{Star} & \colhead{$B$} & 
\colhead{$\sigma_B$} & \colhead{$V$} & 
\colhead{$\sigma_V$} & \colhead{$R$}& \colhead{$\sigma_R$}& \colhead{$I$}& \colhead{$\sigma_I$}}
\startdata 
2004ao-1&  15.37& 0.02&  14.71& 0.03&  14.31& 0.01 & 13.95 &0.01\\
2004ao-2&  15.41& 0.03&  14.45& 0.02&  13.90& 0.01 & 13.37 &0.02\\
2004ao-3&  16.12& 0.05&  15.39& 0.03&  14.99& 0.03 & 14.58 &0.03\\
2004ao-4&  16.83& 0.07&  16.01& 0.04&  15.54& 0.03 & 15.07 &0.04\\
2004ao-5&  17.40& 0.12&  16.71& 0.06&  16.30& 0.05 & 15.92 &0.06\\
2004ao-6&  18.74& 0.23&  17.21& 0.11&  16.02& 0.05 & 14.65 &0.05\\
2004gk-1&  14.81& 0.03&  14.10& 0.03&  13.72& 0.04 & 13.34 &0.03\\
2004gk-2&  16.10& 0.04&  15.43& 0.04&  15.06& 0.02 & 14.62 &0.04\\
2006gi-1&  13.40& 0.01&  12.57& 0.02&  12.13& 0.01 & 11.75 &0.01\\
2006gi-2&  16.59& 0.02&  16.02& 0.03&  15.67& 0.02 & 15.37 &0.04\\
2006gi-3&  17.28 &0.05&  16.43& 0.04&  15.95& 0.03 & 15.45 &0.04\\
\enddata
\label{}
\end{deluxetable}
%%%%%%%%%%%%%%%%%%%%%%%%%%%%%%%%%%%%

The photometric observations were carried out mostly with the 
70-cm reflector of the Sternberg Astronomical Institute in Moscow (M70)
using Apogee AP-47p (a) and AP-7p (b) CCD cameras, and also with the
60-cm reflector of the Crimean 
Observatory of the Sternberg Astronomical Institute (C60)
and 38-cm reflector of the Crimean Astrophysical Observatory (C38), both
equipped with  Apogee AP-47p cameras.

All reductions and photometry were made using IRAF\footnote{IRAF is 
distributed by the National Optical Astronomy Observatory, which is operated 
by AURA under cooperative agreement with the National Science Foundation}.
Photometric measurements of the SNe were made relative to 
local standard stars using PSF-fitting with the IRAF DAOPHOT package.
The background of the host galaxy was negligible for SN 2006gi, which occurred
very far from the center of NGC$~$3147. The projected distance of this
SN from the center of the galaxy is 31.1 kpc, while the radius of the
galaxy is 24.7 kpc (de Vaucouleurs et al. 1991; as given in the 
 RC3 Catalogue of Bright Galaxies and Nasa/Ipac Extragalactic Database).
SN 2004ao was superimposed on a faint spiral arm of UGC$~$10862, and
SN 2004gk was located near the ridge line of an edge-on
spiral galaxy IC$~$3311. For these two objects we applied subtraction of
galaxy background before PSF photometry. We used 
SDSS\footnote{http://das.sdss.org} 
images of the fields for subtraction,
because no high-quality images of these galaxies could be obtained
at our telescopes after the SNe faded.

The passbands of the SDSS photometric system are different from
the response curves of our instruments, and we studied the possible
effect of this fact on the results of our photometry.
We carried out the photometry with galaxy subtraction and without subtraction
and compared the results. On most of the dates the differences
are negligible, and only for few later dates they slightly
exceed $3\sigma$, where $\sigma$ is the uncertainty of photometry as
reported in Tables 3, 4. We also carefully checked the quality of
subtraction and found that it is
quite clear, with no sign of underlying features in the galaxies.
The examination of SDSS images of the host galaxies showed that there
are no bright features at the positions of SNe that could have any
effect on the photometry. We conclude that in the cases of SNe 2004ao and
2004gk the use of
SDSS frames for galaxy subtraction cannot result in significant
errors, comparable to the uncertainties of our photometry.

The $V-$images of SNe with local standard stars are shown in Fig. 1.
The magnitudes of local standards were calibrated on 
photometric nights, when we observed standards from Landolt (1992) and
standard regions in the clusters M~67 (Chevalier and Ilovaisky, 1991),
M~92 and NGC~7790\footnote{http://cadcwww.hia.nrc.ca/standards/}.
They are reported in Table 1.
The filters used at all the telescopes were intended to match
Johnson $B,V,I$ and Cousins $R$ passbands.
The instrumental colour terms were derived from observations of standard
stars. They are presented in Table 2, where the number after the codes
for telescope and camera denotes the filter set.

The $B,V,R,I$ optical photometry and their uncertainties for the three 
SNe are reported in Tables 3, 4 and 5. 
%%%%%%%%%%%%%%%%%%%%%%%%%%%%%%%%%%%
\section{Light Curves}
%%%%%%%%%%%%%%%%%%%%%%%%%%%%%%%%%%%
In this section we describe the main parameters of the individual events, 
 and we present their light curves and colour evolution. 
%%%%%%%%%%%%%%%%%%%%%%%%%%%%%
\begin{deluxetable*}{ccccccccl}
\tablewidth{0pt}
\tabletypesize{\scriptsize}
\tablecolumns{9}
\tablecaption{The instrumental color terms}
\tablehead{\colhead{Telescope,CCD,Filters} &\colhead{$K_b$} & \colhead{$\sigma_{Kb}$} &\colhead{$K_v$}& \colhead{$\sigma_{Kv}$} & 
\colhead{$K_r$}& \colhead{$\sigma_{Kr}$} &
\colhead{$K_i$}& \colhead{$\sigma_{Ki}$} 
%\colhead{CCD,filters} & 
%\colhead{} & \colhead{}& \colhead{}& \colhead{}
%\colhead{} & \colhead{}& \colhead{}& \colhead{}
}
\startdata 
M70a1  & -0.18 &0.01 & -0.011 &0.007 & -0.21 &0.02 & -0.45 & 0.03 \\
M70a2  & -0.21 &0.02 & -0.023 &0.008 &  0.09 &0.01 & -0.39 & 0.03 \\
M70b   & -0.14 &0.01 & -0.023 &0.009 & -0.12 &0.02 & -0.38 & 0.02 \\
C38    & -0.24 &0.02 & -0.026 &0.012 &  0.14 &0.02 & -0.33 & 0.03 \\
C60    &  0.01 &0.01 & -0.050 &0.011 & -0.07 &0.01 & -0.38 & 0.02 \\   
\enddata
\label{}
\end{deluxetable*}

%%%%%%%%%%%%%%%%%%%%%%%%%%
\subsection{\bf{SN 2004ao}}
%%%%%%%%%%%%%%%%%%%%%%%%%%
SN 2004ao was discovered on March 7.54 by the Lick Observatory Supernova Search
(Singer \& Li, 2004; IAUC 8299), with an 
unfiltered  magnitude of 14.9. The supernova was located at
$\alpha =17$$^{\rm h}$28$^{\rm m}$09$^{\rm s}$.35, $\delta = +$07$^{\circ}$24$
^{\prime}$55$^{\prime \prime}$.5
, and lies close to the southern arm of its host galaxy UGC 10862 
(SBc Galaxy). Matheson, Challis \& Kirshner (2004) reported that an 
optical spectrum\footnote{available at http:
//www.cfa.harvard.edu/supernova/spectra}
obtained on March 14.53 was that of a type Ib SNe soon after maximum. The 
spectrum showed characteristics similar to those of SN Ib 1998dt 8 days past maximum 
(Matheson et al. 2001). A nebular spectrum, taken about 3 months after 
 discovery (on June 8.1) at the $J$-band, shows a P-Cygni feature with 
 absorption component at 10430 \AA. The line has
 been identified as He I 10830 \AA, expected from the type Ib nature of 
 the object (Gomez et al. 2004; IAUC 8430).

%%%%%%%%%%%%%%%%%%%%%%%%%%%%%%%%%%%%%
\begin{figure}
 \includegraphics*[width=8cm,height=8cm]{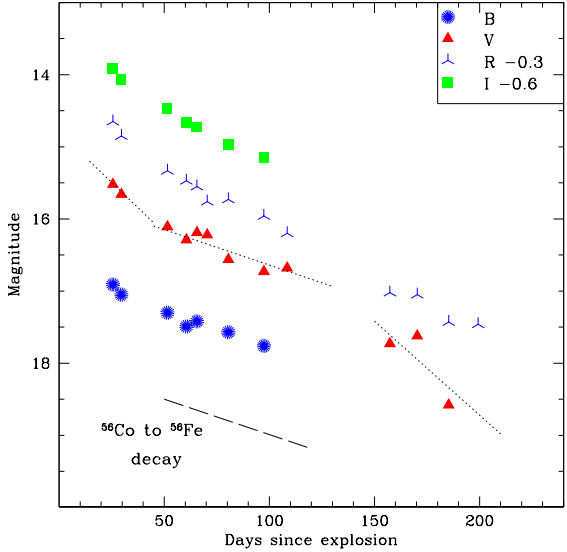}
 \figcaption{$B,V,R,I$ light curves of SN Ib 2004ao. The light curves have 
 been shifted by the reported amounts. Also shown are the fits to the
 three observed phases in the $V$-light curve (dotted lines; see text). The
 $^{56}$Co to $^{56}$Fe decay slope is indicated (long-dashed line)}
 \end{figure}
%%%%%%%%%%%%%%%%%%%%%%%%%%%%%%%%%%%%%

%%%%%%%%%%%%%%%%%%%%%%%%%%%%%%%%%%%%
\begin{deluxetable*}{rccccccccl}
\tablewidth{0pt}
\tabletypesize{\scriptsize}
\tablecolumns{10}
\tablecaption{$BVRI$ photometry of SN 2004ao}
\tablehead{\colhead{JD 2453000+ } & \colhead{$B$} &  \colhead{$\sigma_B$}& \colhead{$V$}&
\colhead{$\sigma_V$} & \colhead{$R$} & \colhead{$\sigma_R$} & \colhead{$I$}& \colhead{$\sigma_I$}& \colhead{Tel.}}
\startdata 
 75.59 &  16.91& 0.06 &  15.52& 0.02 & 14.95& 0.02 & 14.51& 0.03& M70a1\\
 79.60 &  17.05& 0.05 &  15.66& 0.03 & 15.15& 0.03 & 14.67& 0.03& M70a1\\
101.51 &  17.30& 0.06 &  16.11& 0.05 & 15.63& 0.03 & 15.07& 0.04& M70a1\\
110.54 &  17.49& 0.08 &  16.29& 0.03 & 15.77& 0.03 & 15.27& 0.03& M70a1\\
115.54 &  17.42& 0.14 &  16.19& 0.05 & 15.85& 0.06 & 15.32& 0.06& M70a1\\
120.43 &       &      &  16.22& 0.11 & 16.06& 0.15 &      &     & M70a1\\
130.51 &  17.57& 0.13 &  16.56& 0.06 & 16.03& 0.06 & 15.57& 0.06& M70a1\\
147.43 &  17.76& 0.09 &  16.73& 0.07 & 16.26& 0.05 & 15.75& 0.06& M70a1\\
158.45 &       &      &  16.68& 0.05 & 16.50& 0.04 &      &     & M70a1\\
207.36 &       &      &  17.73& 0.08 & 17.32& 0.04 &      &     & M70b\\
220.36 &       &      &  17.62& 0.12 & 17.35& 0.07 &      &     & M70b\\
235.32 &       &      &  18.58& 0.13 & 17.73& 0.05 &      &     & M70b\\
249.31 &       &      &       &      & 17.76& 0.12 &      &     & M70b\\
\enddata
\label{}
\end{deluxetable*}
%%%%%%%%%%%%%%%%%%%%%%%%%%%%%%%%%%

\begin{deluxetable*}{rccccccccl}
\tablewidth{0pt}
\tabletypesize{\scriptsize}
\tablecolumns{10}
\tablecaption{$BVRI$ photometry of SN 2004gk}
\tablehead{\colhead{JD 2453000+ } & \colhead{$B$} &  \colhead{$\sigma_B$}& \colhead{$V$}&
\colhead{$\sigma_V$} & \colhead{$R$} & \colhead{$\sigma_R$} & \colhead{$I$}& \colhead{$\sigma_I$}& \colhead{Tel.}}
\startdata 
341.58 & 15.23& 0.04 & 14.12& 0.03 & 13.84& 0.03&  13.44& 0.04&  C38 \\
389.64 & 16.75& 0.05 & 15.44& 0.03 & 15.14& 0.02&  14.39& 0.02&  M70a2\\
406.41 & 16.91& 0.03 & 15.71& 0.03 & 15.45& 0.04&  14.77& 0.03&  M70a2\\
432.47 & 17.13& 0.04 & 16.13& 0.03 & 15.76& 0.02&  15.25& 0.03&  M70a2\\
446.50 & 17.18& 0.05 & 16.29& 0.03 & 15.91& 0.03&  15.37& 0.03&  M70a2\\
455.50 & 17.46& 0.05 & 16.40& 0.04 & 16.06& 0.02&  15.55& 0.04&  M70a2\\
463.39 & 17.61& 0.06 & 16.51& 0.04 & 16.07& 0.04&  15.64& 0.06&  M70a2\\
468.43 & 17.43& 0.06 & 16.69& 0.04 & 16.20& 0.05&  15.70& 0.05&  M70a2\\
506.39 & 18.15& 0.06 & 17.64& 0.07 & 16.97& 0.04&  16.70& 0.07&  M70b\\
\enddata
\label{}
\end{deluxetable*}
%%%%%%%%%%%%%%%%%%%%%%%%%%%%%%%%%%%%

\begin{deluxetable*}{rccccccccl}
\tablewidth{0pt}
\tabletypesize{\scriptsize}
\tablecolumns{10}
\tablecaption{$BVRI$ photometry of SN 2006gi}
\tablehead{\colhead{JD 2453000+ } & \colhead{$B$} &  \colhead{$\sigma_B$}& \colhead{$V$}&
\colhead{$\sigma_V$} & \colhead{$R$} & \colhead{$\sigma_R$} & \colhead{$I$}& \colhead{$\sigma_I$}& \colhead{Tel.}}
\startdata 
998.67\tablenotemark{a}&17.03 &0.06&  16.30& 0.04&  15.91& 0.02&  15.64& 0.07& NOT\\
999.22  & 17.10& 0.05 & 16.18& 0.05&  15.91& 0.04 & 15.63& 0.05&  M70b \\
1003.54 & 17.52& 0.05 & 16.37& 0.03&  15.97& 0.02 & 15.72& 0.03&  M70b \\
1006.58 & 18.27& 0.12 & 16.69& 0.06&  16.05& 0.04 & 15.80& 0.05&  M70b \\
1026.52 & 18.95& 0.17 & 17.66& 0.14&  16.99& 0.04 &      &     &  M70b \\
1044.44 &      &      & 17.67& 0.07&  17.22& 0.04 &      &     &  C60  \\
1048.62 & 19.15& 0.21 & 17.98& 0.06&  17.35& 0.03 & 16.84& 0.14&  C60  \\
1059.61 & 19.18& 0.04 & 18.02& 0.02&  17.53& 0.04 &      &     &  C60  \\
1060.57 & 19.49& 0.16 & 18.09& 0.05&  17.54& 0.05 &      &     &  C60  \\
1118.43 &      &      & 18.95& 0.16&  18.31& 0.07 & 17.90& 0.11&  M70b \\
1131.34 &      &      & 19.00& 0.40&  18.45& 0.27 &      &     &  M70b \\
\enddata
\tablenotetext{a}{Measured on ALFOSC-camera images (NOT observatory).}
\label{}
\end{deluxetable*}
%%%%%%%%%%%%%%%%%%%%%%%%%%%%%%%%%%%%%

 In Fig. 2, the $B,V,R,I$ light curves are displayed. The early 
 photometric evolution together with the nature 
 of the March 14.53 spectrum (Matheson, Challis \& Kirshner, 2004) provide 
 the possibility of estimating
 the explosion date. In fact, based on our early photometric data, and bearing 
 in mind the typical rise time for type Ib/c SNe, between shock breakout and
 maximum light, to be $15-20$ days, we 
 estimate the explosion time to be January 21, 2004 within an uncertainty of 
 5 days (JD 2453050 $\pm$ 5 d). 
%%%%%%%%%%%%%%%%%%%%%%%%%%%%%%%%%%%%%%%
\begin{figure}
 \includegraphics*[width=8cm,height=8cm]{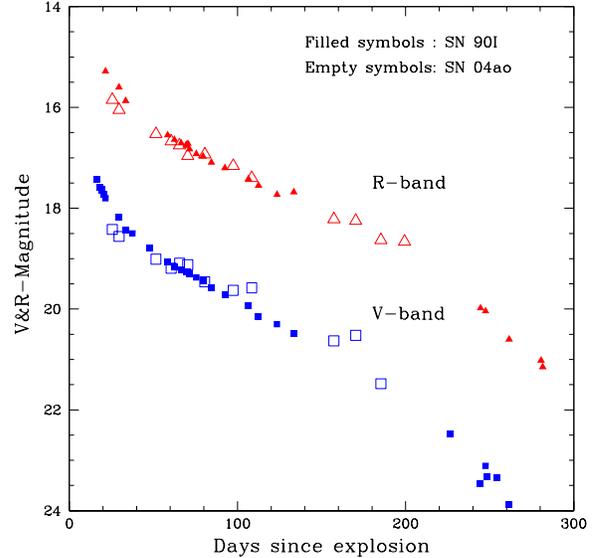}
 \figcaption{Comparison of SN 2004ao to SN Ib 1990I in the $V$ and
 $R$ bands. SN 2004ao data are vertically shifted to fit SN 1990I data.}
 \end{figure}
%%%%%%%%%%%%%%%%%%%%%%%%%%%%%%%%%%%%%%%

 A comparison, in the $V$ and $R$ bands, with 
 the well studied SN Ib 1990I (Elmhamdi et al. 2004)
 is displayed in Fig. 3, where the SN 2004ao data are vertically shifted 
 to fit SN 1990I data. The match in the time range [25-110] days is good. Later
 , SN 2004ao seems to evolve slightly less steeply than SN 1990I, 
 although the lack of late 
 photometric observations in SN 2004ao, after day 200, inhibits a 
 complete comparison between the two events. 
 The good early match however provides support for 
 our explosion time estimate. Our first photometric observations occur then
 at the end of the post-maximum decline phase.   
 
 SN 2004ao suffers high galactic extinction, $A_V^{gal}=0.348$ mag, according
 to maps of the galactic dust distribution by Schlegel, Finkbeiner $\&$ Davis 
 (1998). This corresponds to a colour excess 
 of about $E(B-V)=0.105$, where the standard reddening laws of Cardelli et al. 
 (1989) have been used. The available early spectra of the object
 do not display an obvious interstellar Na I D absorption at the velocity 
 of the host galaxy, however an upper limit to the line equivalent width (EW)
 of 0.07\AA$~$ is estimated (M. Modjaz, Private communication).
 Adopting the general  correlation between 
  Na I D EW and colour excess, derived by Turatto et al. (2003), the total 
 extinction suffered by SN 2004ao is then $A_V^{total}=0.382$ mag.

 The radial velocity of the host
 galaxy with respect to the CMB radiation is 1668 km s$^{-1}$, while the
 one corrected for the Local Group infall onto the Virgo Cluster is 
 1813 km s$^{-1}$ (Theureau G. et al. 1998; Paturel et al. 2003;
 as given by 
 LEDA\footnote{http://leda.univ-lyon1.fr} extragalactic database).
 This later value translates into a distance modulus $\mu \sim 32.06$
 (using H$_0=$70 km s$^{-1}$Mpc$^{-1}$).

%%%%%%%%%%%%%%%%%%%%%%%%%%%%%%%%%%%%%
\subsection{\bf{SN 2004gk}}
%%%%%%%%%%%%%%%%%%%%%%%%%%%%%%%%%%%%%
\begin{figure}
 \includegraphics*[width=8cm,height=8cm]{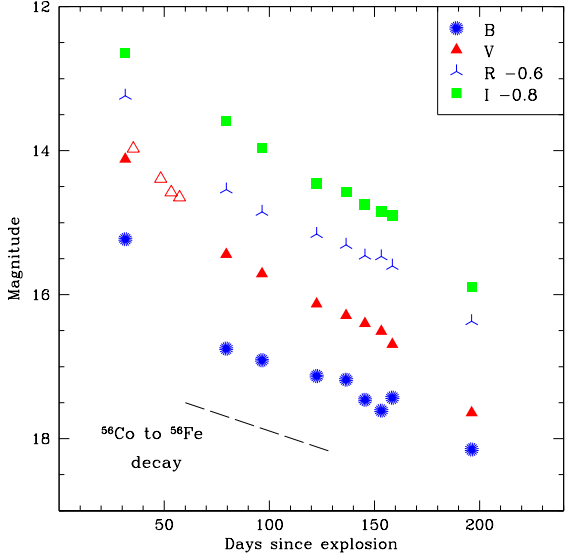}
 \figcaption{$B,V,R,I$ light curves of SN Ic 2004gk. The light curves have 
 been shifted by the reported amounts. Some early V-band data, plotted in 
 empty triangles, are taken from ``Astrosurf'' amateurs 
 astronomer service.}
 \end{figure}
%%%%%%%%%%%%%%%%%%%%%%%%%%%%%%%%%%%%%%%

The discovery of SN 2004gk was reported on
November 25.5 by R. Quimby et al. 2004 (IAUC 8446). The SN is located at 
$\alpha =12^{\rm h}25^{\rm m}33^{\rm s}.21$, $\delta =+12^{\circ}15^{\prime}
39^{\prime \prime}.9$ (equinox 2000.0), corresponding to $1^{\prime \prime}.5$ 
west and $2^{\prime \prime}.7$ north of 
the center of the host Sc galaxy IC 3311. An image taken on June 12.17, with
the McDonald Observatory 0.45-m ROTSE-IIIb telescope, shows nothing at the
 supernova position, limiting its magnitude to about 17.9 (IAUC 8446). The 
 object was classified as a type Ic supernova near maximum. Indeed the 
optical spectrum obtained on November 27.52 by M. Shetrone $\&$ V. Riley with 
 the 9.2-m Hobby/Eberly telescope resembles that of the intensively 
 studied Ic SN 1994I
 near maximum light (IAUC 8446). Moreover, the radio observation of 
 SN 2004gk on November 30.5 indicates a comparable radio luminosity ( 
 $\sim 10^{25}$erg/s/Hz) to that of SN Ic 2002ap at similar epoch (Soderberg 
 et al. 2004). A nebular spectrum was reported by 
 Modjaz et al.(2008). It also shows a clear a type Ic nebular behaviour. 

 The $B,V,R,I$ light curves are illustrated in Fig. 4. The Figure 
 includes also four early $V$-band estimates from ``Astrosurf'' amateur 
 astronomers service\footnote{http://www.astrosurf.com/snweb2/index.html} 
 (plotted as empty triangles). According to the evolution of the 
 observed early photometry and the nature of the November 27.52 spectrum 
 (IAUC 8446), taking into account
 as well the typical rise time in type Ib/c SNe, it is possible to constrain
 the explosion date. We estimate it to be November 2, 2004 
 (JD 2453312) and with an uncertainty of 5 days. An alternative method
 supporting our estimated date is by comparing SN 2004gk with the well 
 studied type Ic SN 1994I. Fig. 5 compares the $V$ and $R$ light curves of 
  the two events. SN 2004gk data are vertically displaced to fit SN 1994I
  data. The goodness of the match indicates the epoch of our first photometric 
 observations to be near the end of the post-maximum decline phase. 
 Data around day 120 and later seem to indicate a less steep decline rate 
 in SN 2004gk than in SN 1994I, although the data point around day 195 
 shows a similar evolution trend. 

 The galactic extinction toward the host IC 3311, 
 according to maps of the galactic dust distribution by 
 Schlegel et al. (1998), is $A_V^{gal}=0.1$ mag.
 The standard reddening laws of Cardelli et al.(1989) are again adopted.
 Additional intra-galaxy reddening is
  supported by the presence in the spectra of narrow Na I D line at the 
  host galaxy redshift with an equivalent width of EW$=$0.75$\AA$ (Robert 
  Quimby; Private communication). Adopting the general  correlation between 
  Na I D EW and colour excess, derived by Turatto et al. (2003), we estimate
   $A_V^{host}=0.372$ mag. The total extinction suffered by SN 2004gk is then
   $A_V^{total}=0.472$ mag.
%%%%%%%%%%%%%%%%%%%%%%%%%%%%%%%%%%%%%%%
\begin{figure}[h]
 \includegraphics*[width=8cm,height=8cm]{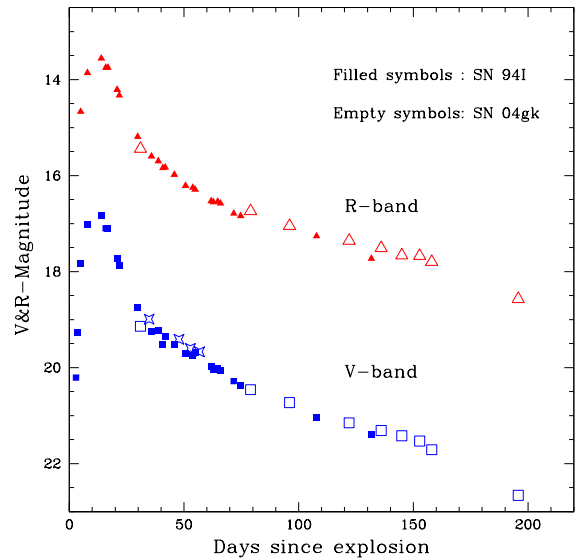}
 \figcaption{$V$ and $R$ bands light curves of SN 2004gk compared to SN Ic 
 1994I. SN 2004gk data are vertically shifted to fit SN 1994I data.}
 \end{figure}
%%%%%%%%%%%%%%%%%%%%%%%%%%%%%%%%%%%%%%%

 The radial velocity of the host
 galaxy corrected for the Local Group infall onto the Virgo Cluster is
 -79 km s$^{-1}$ while the one with respect to the CMB radiation
 is 160 km s$^{-1}$ (Springob C. M. et al. 2005; Paturel et al. 2003; 
 as reported in the LEDA database). These small redshifts
 are unreliable indicators of Huble flow and their corresponding distances 
 are unusable. However since IC 3311 is found to be close to the center of 
 the Virgo cluster one can then adopt the mean distance of 
 the cluster. An accurate recent value has
 been  found by Mei et al. (2007), to be 16.5 Mpc, adopted in our paper.

%%%%%%%%%%%%%%%%%%%%%%%%%%%%%%%%%%%%%%%
\subsection{\bf{SN 2006gi}}
%%%%%%%%%%%%%%%%%%%%%%%%%%%%%%%%%%%%%%%
%%%%%%%%%%%%%%%%%%%%%%%%%%%%%%%%%%%%%%%
\begin{figure}
 \includegraphics*[width=8cm,height=8cm]{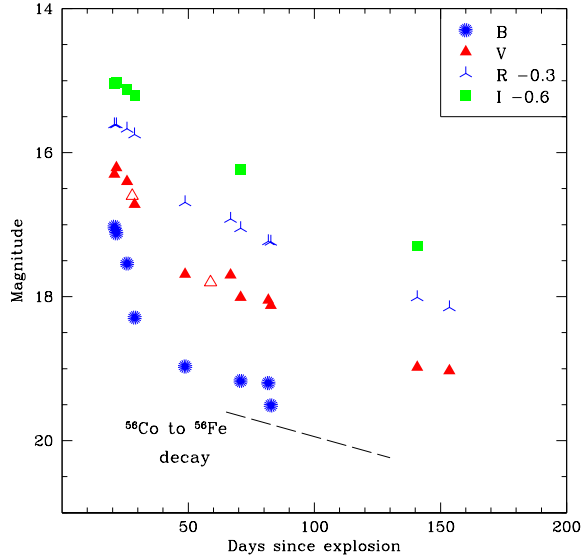}
 \figcaption{$B,V,R,I$ light curves of SN Ib 2006gi. The light curves have 
 been shifted by the reported amounts. Additional V-band data, shown in 
 empty triangles, are taken from ``Astrosurf'' amateurs 
 astronomer service.}
 \end{figure}
%%%%%%%%%%%%%%%%%%%%%%%%%%%%%%%%%%%%%%%

SN 2006gi was discovered on September 18.8 by K. Itagaki with an unfiltered
 magnitude of 16.3 (IAUC 8751). With coordinates of 
$\alpha =10^{\rm h}16^{\rm m}46^{\rm s}.76$, $\delta =+73^{\circ}26^{\prime}26
^{\prime \prime}.4$ (equinox 2000.0). SN 2006gi was located $30^{\prime \prime}
$ west and $144^{\prime \prime}$ north of 
the center of the host Sbc galaxy NGC 3147. 
Nothing was seen at the location of
the supernova on March 22 and June 3 images (limiting magnitude 19.0; 
Nakano $\&$ Itagaki 2006). Moreover, a {\bf{NOT}} (Nordic Optical Telescope) 
spectrum taken on September 19 displays 
 similarities with the well studied type Ib SNe 1984L and 1990I 
 around maximum light\footnote{See 
 http://www.supernovae.net/sn2006/sn2006gi.pdf}, indicating the type Ib 
 nature of the event (Stanishev et al. 2006). 
 Four SNe have been already detected in the host NGC 3147, namely SNe 1972H,
 1997bq, 2006gi and the very recent event 2008fv.

 Fig. 6 presents the $B,V,R,I$ light curves of SN 2006gi. Two additional 
 $V-$band photometric points from ``Astrosurf'' amateur astronomers service 
 are added in the figure. On the one hand the shape of the early 
 photometry indicates the observations to occur near maximum light. 
 On the other hand, the features 
 in the {\bf{NOT}} spectrum together with the discovery date
 suggest constraints on the age of the SN. We indeed estimate the 
 explosion time to be August 20, 2006 (JD 2453978) with an uncertainty of 
 5 days. In Fig. 7 SN 2006gi is compared with the type Ib SNe 1999ex 
 (Hamuy et al. 2002) and 1999dn (Benetti et al. 2010)
 , where SNe data are vertically displaced to fit together. 
  Our first observations correspond to the near post-maximum declining stage.
 The good match supports our estimate of the explosion date.
 Note in addition the good fit with SN 1999dn data at late times.

  The galactic extinction due to the Milky Way toward the parent 
 galaxy of SN 2006gi, 
 according again to maps of the galactic dust distribution by 
 Schlegel et al. (1998), is $A_V^{gal}=0.077$ mag, the 
 standard reddening laws of Cardelli et al. (1989) again being adopted.
 Furthermore, the {\bf{NOT}} spectrum shows the presence of narrow Na I D 
 line at the host galaxy redshift with an equivalent width of 
  EW$=$0.6$\AA$ (V. Stanishev, Private commmunication). The general 
  correlation by Turatto et al. (2003) is then
  used to translate this EW to reddening. The total extinction suffered 
  by SN 2006gi is estimated to be $A_V^{total}=0.38$ mag.
 
 The recession velocity of the host 
 galaxy NGC 3147 relative to the CMB radiation and the one corrected for 
 the Local Group infall onto the Virgo Cluster respectively are
 2875 km s$^{-1}$ and 3077 km s$^{-1}$ (Saunders W. et al. 2000; 
 Springob C. M. et al. 2005; 
 Paturel et al. 2003; as given in the the LEDA extragalactic database). 
 This latter translates into a distance 
 modulus $\mu \sim 33.2$.

 Hereafter, throughout the paper, we will adopt the above constrained  
 parameters for the three events (i.e. explosion date, total extinction 
 and distance modulus). 
%%%%%%%%%%%%%%%%%%%%%%%%%%%%%%%%%%%%%%%
\begin{figure}
 \includegraphics*[width=8cm,height=8cm]{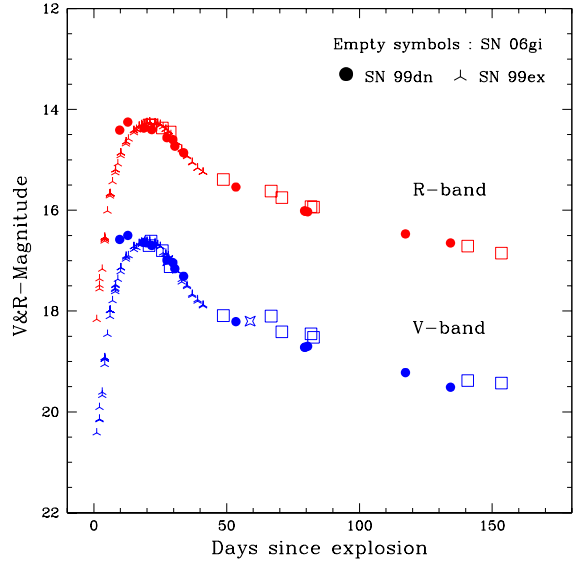}
 \figcaption{Comparison of the $V$ and $R$ light curves of SN 2006gi with SNe 
 Ib 1999dn and 1999ex. The data are arbitrary vertically shifted 
 to fit together}
 \end{figure}
%%%%%%%%%%%%%%%%%%%%%%%%%%%%%%%%%%%%%%%

%%%%%%%%%%%%%%%%%%%%%%%%%%%%%%%%%%%%
\section{Photometric evolution}
%%%%%%%%%%%%%%%%%%%%%%%%%%%%%%%%%%%%%%%
%%%%%%%%%%%%%%%%%%%%%%%%%%%%%%%%%%%%%%%
\begin{figure}
 \includegraphics*[width=9cm,height=8.5cm]{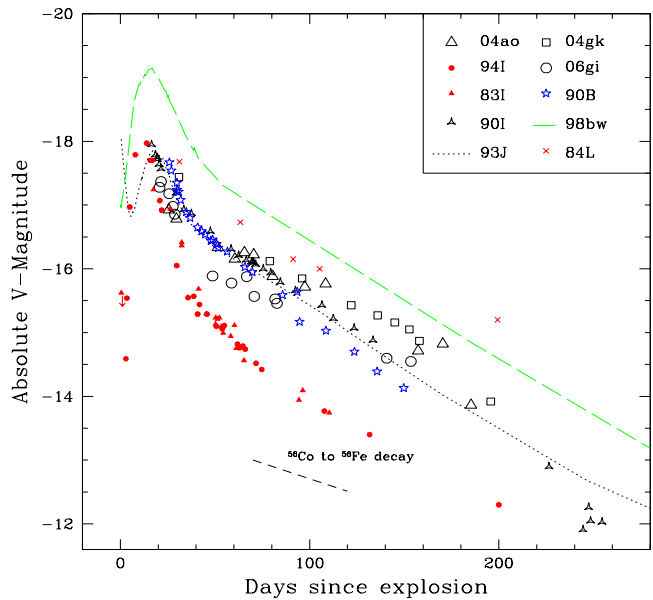}
 \figcaption{Comparison of the $V$ absolute light curves of SNe 2004ao,
 2004gk and 2006gi with those of other Ib/c events. The
 $^{56}$Co to $^{56}$Fe decay slope is shown (short-dashed line)}
 \end{figure}
%%%%%%%%%%%%%%%%%%%%%%%%%%%%%%%%%%%%%%%

 In the following analysis we discuss the nature of the light 
 curves and their different evolutionary phases,
 comparing our three SNe Ib/c with other well studied events.\\

 In Fig. 2 three possible different phases are
 recognizable for SN 2004ao, with approximate time intervals 
 [0-50] days, [50-150] days and later than 150 days.
 The corresponding decline rates are 
 indicated by the dotted lines. The exponential life-time of $^{56}$Co , 
 111.26 days or 0.976  mag per 100 days, 
 is also shown for comparison.
 Similar phases are observed in both SN 2004gk (Fig. 4) and SN 2006gi (Fig. 6).
  The weighted linear least-squares fits to the $B,V,R,I$ observations 
 and their errors for the three events at 
 the ``possible three'' phases are summarized in Table 6. 
 
 Clearly, the three different
 decline stages appear more prominent in SN 2004ao compared to SNe 2004gk 
 and 2006gi. This may be due to the better temporal coverage
 for SN 2004ao. In addition, the phases become clearer and better resolved
  as we progress toward the blue.
  As can be seen from Fig.6 and Table 6, the
  post-maximum decline rate is higher in SN 2006gi, ($\gamma_B\sim$
  6.2 mag (100 d)$^{-1}$), and for both SNe 2004gk and 2006gi seems to 
  be much faster in the blue bands than in the red ones, while in SN 2004ao
  it increases toward the red bands. During this early phase,
  SN 2006gi resembles SNe Ib 1990I (Elmhamdi et al. 2004) and 1999ex 
 (Hamuy et al. 2002) in the large decline rate with a decrease toward 
 the red bands. We note here that our first
 photometric data for SN 2004gk occur at the tail part of the early post
 maximum decline phase (see comparison with SN Ic 1994I). This means that
 the computed early decline is under-estimated. Indeed most type Ic
 SNe seem to have a much faster decline rate at this phase than SNe Ib 
 ($\gamma_V\sim$ 11 mag (100 d)$^{-1}$ and $\sim$ 8.5 mag (100 d)$^{-1}$
  for the well observed SNe Ic 1994I and 1983I, respectively; 
  Elmhamdi et al. 2004).
 %%%%%%%%%%%%%%%%%%%%%%%%%%%%%%%%%%%%%
\begin{deluxetable*}{llllll}
\tablewidth{0pt}
\tabletypesize{\scriptsize}
\tablecolumns{6}
\tablecaption{Decline rates of the 3 events.}
\tablehead{\colhead{} &\colhead{Supernova}& \colhead{$\gamma_B$} & \colhead{$\gamma_V$}& \colhead{$\gamma_R$}&
\colhead{$\gamma_I$} \\ \colhead{} & \colhead{name}&\colhead{mag (100 d)$^{-1}$}&\colhead{mag (100 d)$^{-1}$}
&\colhead{mag (100 d)$^{-1}$}&\colhead{mag (100 d)$^{-1}$}}
\startdata 
%\hline
 &SN04ao &2.5 &2.9 &3.6 & 3.5 \\
Phase 1 &SN04gk &3.1 &2.8 &2.7 &2\\
& SN06gi&6.2 &5.4 &4.1 & 2.3 \\
\hline
 &SN04ao &0.91 &1.02 &1.37 &1.47  \\
Phase 2 &SN04gk\tablenotemark{a}&1.02 &1.5 &1.27 &1.71 \\
&SN06gi &1.1 &1.26 &1.46 &1.51  \\
\hline
&SN04ao &$--$ & 2.6&1.32 & $--$ \\
Phase 3 &SN04gk\tablenotemark{b} &1.49 &2.58 &2.07 &2.52 \\
&SN06gi &$--$ &$--$ &$--$ &$--$  \\
\enddata
\tablenotetext{a}{The first and last data are excluded.}
\tablenotetext{b}{Using only the last three points}.
\label{}
\end{deluxetable*}
%%%%%%%%%%%%%%%%%%%%%%%%%%%%%%%%%%%%%

%%%%%%%%%%%%%%%%%%%%%%%%%%%%%%%%%%%%%%%
  During the second phase SN 2006gi is found to be much faster 
  in the red bands than
  in the blue ones, opposite to what is seen during the first phase. This 
  rate increase trend toward the red is also observed in SN 2004ao.
  SN 2004ao has a $V-$band slope very close to the full $\gamma$-ray 
  trapping rate. This may be of interest since this
  behaviour characterizes type II SNe, which have higher ejecta masses, 
  rather than Ib/c events, but is also reminiscent of what was seen in the 
  peculiar type Ib SN 1984L (Schlegel $\&$ Kirshner, 1989) although their 
  behaviour differs at later phases.

%%%%%%%%%%%%%%%%%%%%%%%%%%%%%%%%%%%%%%%
\begin{figure}
 \includegraphics*[width=8cm,height=8cm]{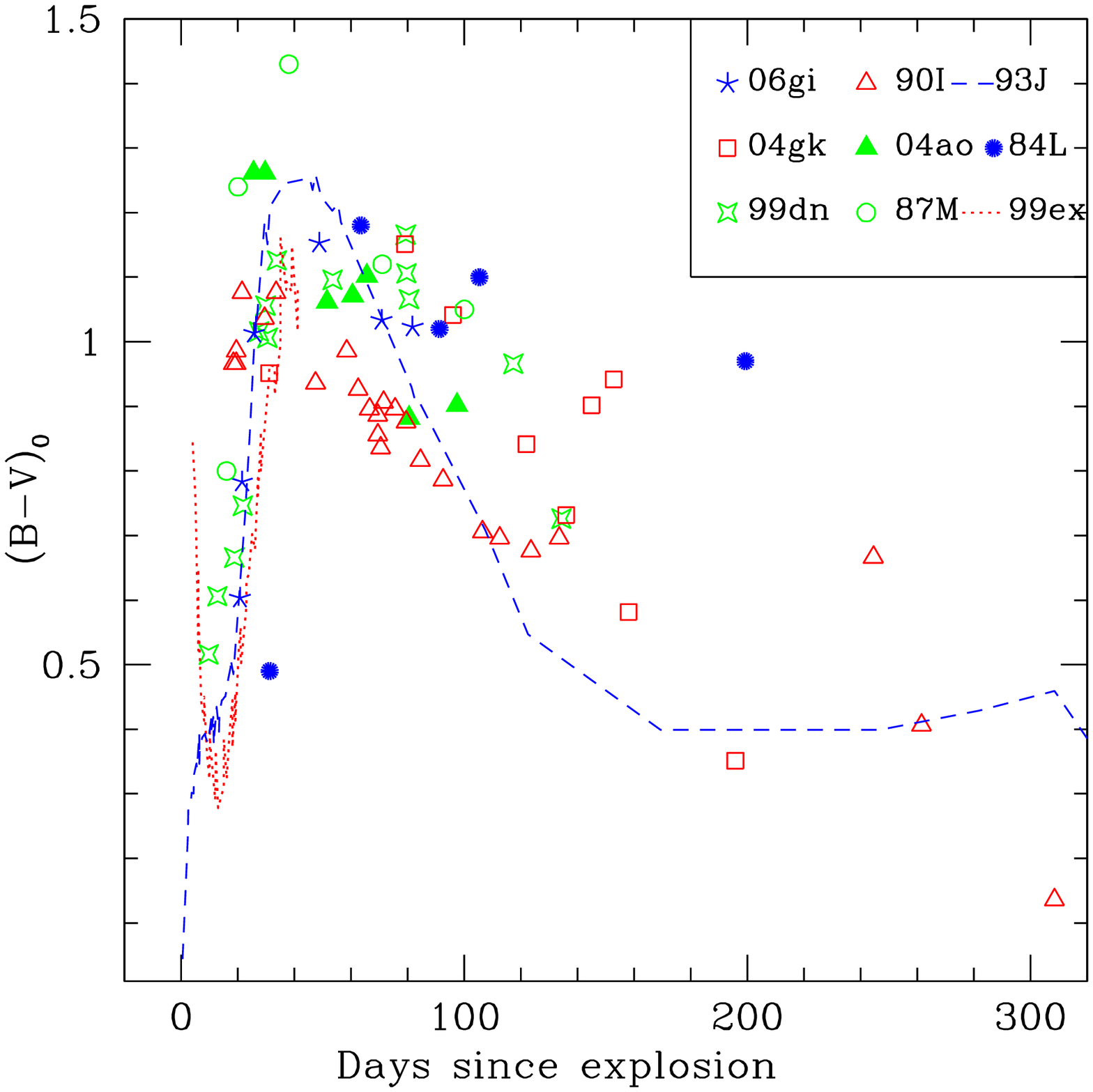}
 \figcaption{Comparison of the evolution of the $B-V$ intrinsic colour of 
SNe 2004ao, 2004gk and 2006gi with other Ib/c events.}
 \end{figure}
%%%%%%%%%%%%%%%%%%%%%%%%%%%%%%%%%%%%%%%
  In Figure 8 the absolute $V$ 
  light curves of SNe 2004ao, 2004gk and 2006gi are compared with other type 
  Ib/c core collapse events. Around day 100 since explosion SN Ib 1984L appears
  $\sim$0.3 mag brighter than SN 2004ao, whereas the difference 
  becomes $\sim$1 mag at day 200.  
  The SN 2004gk photometric point around day 200 reflects a deviation, 
  possibly occurring after day 150, from a second phase slope trend . 
  Figure 9 reports the $B-V$ intrinsic colour of our three
  SNe together with that of other type Ib/c SNe for comparison. $(B-V)_0$ 
  of SN IIb 1993J is also shown (dashed line). The figure highlights
  the similar evolutionary trend followed by the sample events, displaying 
  a rapid reddening during the first 50 days, reflecting the cooling 
  due to the envelope expansion. After the peak the SNe turn to 
  the blue with almost similar slopes. 
  Then the events settle on
  a less varying colour phase. Later on SNe 1990I and 2004gk
  show a sudden drop, with different epochs of occurrence. 
  While SN 1990I colour falls around day 250, SN 2004gk exhibits a
  steep decline at day 150 after explosion, with a rate of 
  $\sim$0.6 mag in 50 days.

%%%%%%%%%%%%%%%%%%%%%%%%%%%%%%%%%%%%%%%%%%%%%%%%%%%%%%%%%%%%%%%%%
\section{The Bolometric luminosities and the physical parameters}
%%%%%%%%%%%%%%%%%%%%%%%%%%%%%%%%%%%%%%%%%%%%%%%%%%%%%%%%%%%%%%%%

 The fundamental physics of SNe is imprinted on the bolometric 
 luminosities. Indeed, in order to obtain an accurate estimate of
 the physical parameters 
 one needs to treat and model the bolometric light curves rather than 
 only analyzing the individual broad band photometry evolution, especially for
 cases where the optical light curves do not track the bolometric ones. 
 Unfortunately, to date only a few core collapse SNe have good 
 coverage with broad-band 
 photometry (spanning the X-ray to the IR regions of the electromagnetic 
 spectrum), thus limiting  a complete and correct understanding
 of how physically these classes of events behave and evolve.  
    
%%%%%%%%%%%%%%%%%%%%%%%%%%%%%%%%%%%%%
\subsection{Bolometric light curves}
%%%%%%%%%%%%%%%%%%%%%%%%%%%%%%%%%%%%%
     
 Two methods are usually adopted to construct the bolometric luminosities.
 The first uses a SED, spectral energy distribution, analysis in deriving 
 the equivalent corresponding 
  BB luminosities, while in the second method the observed broad band 
 magnitudes are directly integrated. 

 In the present work we adopt the second method. Our broad band observations
 are corrected for extinction effects adopting the standard reddening 
 curve of Cardelli et al. (1989). The magnitudes are
 converted then into monochromatic fluxes.
 The distance to the SN host galaxy is used to convert
 the summed fluxes to luminosities.
 We note here that when for a given passband a night is lacking we 
 extend the corresponding data
 using interpolation or extrapolation of the existing data nights adopting 
 low order polynomial fits.
      
%%%%%%%%%%%%%%%%%%%%%%%%%%%%%%%%%%%%%%%
\begin{figure}
 \includegraphics*[width=9.5cm,height=9cm]{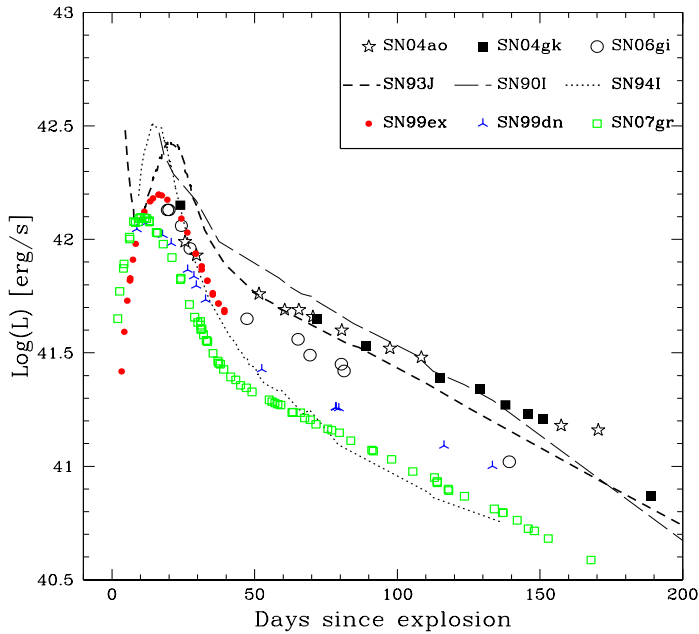}
 \figcaption{Comparison of the pseudo-bolometric light curves of SNe 2004ao, 2004gk and 2006gi with those constructed for other Ib/c events.}
 \end{figure}
%%%%%%%%%%%%%%%%%%%%%%%%%%%%%%%%%%%%%%%
 Moreover, the derived $BVRI$ luminosities should include 
 the near-IR. At present few Ib/c objects were 
 photometrically followed in the IR. In the case of SN 1998bw, a near-IR 
 fraction of 42$\%$ at day 370 was estimated to be added to 
 the $L_{BVRI}$ (Sollerman et al. 2002), whereas Patat et al. (2001) have 
  derived a fraction of about 35$\%$ on day 65.4. Taubenberger et al. (2006)
 found that, for SN Ic 2004aw, the near-IR contribution increases 
 from 31$\%$ to 45$\%$ between
 day 10 and day 30 past B-maximum light. Recently, SN Ib 2008D was found 
  to have a near-IR contribution about 24$\%$ at age 30 days since outburst 
 (Modjaz et al. 2009).
  Though not precise, in view of the possible 
  increasing importance of the near-IR fraction with time (e.g. SN 1998bw; 
  Sollerman et al. 2002), here we simply adopt 35$\%$ as a constant 
  scaling factor due to the contribution of the near-IR fluxes. 

  The resulting quasi-bolometric light curves for the 3 SNe are displayed
 in Fig. 10, and compared with those of other stripped-envelope events.
 These include: SN Ib 1990I (Elmhamdi et al. 2004), SN Ic 1994I (Richmond et 
 al. 1996a), SN IIb 1993J (Barbon et al. 1995; Richmond et al. 1996b),
 SN Ib 1999ex (Stritzinger et al. 2002), SN Ib 1999dn (Benetti et al. 2010)
 and SN Ic 2007gr (Valenti et al. 2008; Hunter et al. 2009). The reported
 quasi-bolmetric light curves were constructed following the same steps as
 explained earlier (i.e. for SNe 2004ao, 2004gk and 2006gi), and adopting
 extinction values and distance moduli from the cited corresponding 
 reference papers. 

 SNe 2004ao, 2004gk, 1993J appear to 
 have similar luminosities over the time interval [50-100] days. After
 day 100, both SNe 2004ao and 2004gk evolve slower, i.e greater $e-$folding
 time, compared to SNe 
 1990I and 1993J. At this phase, SN 2006gi is found to be $\sim$0.15 dex 
 fainter than SNe 2004ao and 2004gk and $\sim$0.15 dex brigther than SN 1999dn.
 Interestingly, SN 2006gi is similar to
 SN 1999ex at earlier epoch, in the interval [10-50] days, 
 with almost identical luminosities and decline rate. Furthermore,
 although SN 1999dn appears fainter than SN 2006gi, their observed decline
 rates indicate an overall shape similarity in the interval [15-100] days.
 Around day 140, the luminosity difference seems decreasing between the 
 two events. 
  Later on, the derived luminosity of SN 2004ao 
 around day 170 is $\sim$0.2 dex brighter than SNe 1990I and 1993J. 
 Around day 190, SN 2004gk seems to change its previous trend converging to 
 a comparable luminosity as SN 1993J. 
 SNe 1994I and 2007gr are the objects with the lowest
 luminosities over the late evolutionary phase.
   
%%%%%%%%%%%%%%%%%%%%%%%%%%%%%%%%%%%%%
\subsection{Physical parameters}
%%%%%%%%%%%%%%%%%%%%%%%%%%%%%%%%%%%%%
  After constructing the bolometric light curves, we adopt a simple model 
 for radioactive powering especially during the radioactive tail after the 
 steep decline from the maximum brightness. A reasonable fit to the data
 provides quantitative insights on the ejecta and $^{56}$Ni masses
 being the basic parameters responsible for the bolometric light curve shape.

 Our principal goal is to model-fit the 
 bolometric behaviour starting at the transition phase, i.e. about 50 days up
 to about 150 days. In this time
 range we ignore the contributions from radioactive elements other 
 than $^{56}$Ni and $^{56}$Co. Longer lived isotopes would have little effect.

  The simple model describes the luminosity evolution in an homologously 
  expanding 
  spherical ejecta with a point source $\gamma$-ray deposition from 
  $^{56}$Ni$\rightarrow$$^{56}$Co$\rightarrow$$^{56}$Fe. It is based on the 
  simple approach adopted by Swartz $\&$ Wheeler (1991) and Clocchiatti $\&$ 
  Wheeler (1997). By assuming free expansion of the ejecta ($v(r,t) = r/t$), a 
  power law density distribution ($\rho (r,t) \propto r^{-n}(t)$) and a
  $\gamma$-ray opacity ($\kappa_\gamma$) constant throughout the ejecta, the 
  equations for the mass, kinetic energy and $\gamma$-ray optical depth can be 
  solved. The total $\gamma$-ray optical 
  depth can be expressed then in terms of the main physical parameters 
  (Clocchiatti $\&$ Wheeler 1997): 
  \begin{equation}
  \tau _\gamma = C \kappa_\gamma \times M_{ej}^2/E_{K} \times t^{-2}
  \end{equation}

  Where t refers to time since explosion. For a power index $n=7$ we 
  have $C\simeq 0.053$, adopted in our calculations.

  With these simplified considerations, the emergent luminosity attenuated 
  by the ejecta, i.e thermalized by the ejecta, at a given time t, reads:
   \begin{equation}
   L(t) = L_0(t) \times [1 - exp(-\tau _\gamma(t))]
   \end{equation}
    L$_0(t)$ is the $\gamma$-rays luminosity from radioactive decay.

 L$_0$ at a given time is estimated from the radioactive 
 energy properties of the $^{56}$Ni$\rightarrow$$^{56}$Co$\rightarrow$$^{56}$Fe
  decay, adopting the main parameters presented by Jeffery (1999) and 
 Nadyozhin (1994). Once the total rate of radioactive energy production 
 at a given time is computed we then combine it with equations 1 and 2. This 
 provides a general description of the simple radioactive decay energy 
 deposition model in a spherical geometry.  

 A consistent light curve model should
 consider simultaneously details from the bolometric light curve and the 
 velocities. To break the degeneracy between $E_{K}$ and $M_{ej}$, we rely
 on the available information about the velocity of the events.
 In a recent work, Maurer et al. (2010) have studied the velocity of a sample 
 of 56 stripped-envelope supernovae by means of their nebular spectra.
  The sample includes both SN 2004ao and SN 2004gk. 
 The authors report two characteristic velocities for the whole sample, 
 namely $v_\alpha$ (related to the ratio of core kinetic energy to core mass 
 as estimated from spectral modeling; equations 2 and 3 in Maurer et al. 2010) 
 and $v_{50}$ (directly measured from the half width of the oxygen doublet 
 [O I] 6300,6364 \AA$~$at half maximum intensity). 
 Both velocities, i.e. $v_\alpha$ and $v_{50}$, 
 characterize the inner ejecta of the sample events. For most SNe of
 a given class, the study revealed a quit similar average core velocities: 
 5126 $\pm$816 km s$^{-1}$ for 27 Ic SNe; 4844 $\pm$935 km s$^{-1}$ 
 for 13 Ib SNe and 4402 $\pm$403 km s$^{-1}$ for SNe of type IIb.    
 The energetic subclass, termed broad-lined SNe (BL-SNe or hypernovae; 
 12 objects in the studied CCSNe sample), show 
 on average a higher velocity of 5685 $\pm$842 km s$^{-1}$. One of the   
 most important results of the cited paper is that not all the BL objects
 behave in the same way as they evolve in time. In fact, while at early epochs
 the BL-SNe seem to be homogeneous in having their outer ejecta moving at 
 very high velocities, the core velocities show a large variations at late
 phases. The asphericity of the inner ejecta of the BL SNe class might be
 an explanation. 
   
 As far as SNe 2004ao and 2004gk are concerned, they appear to belong to 
 the regular velocity behaviour classes (see table 3 in Maurer et al. 2010). 

Therefore, we find it reasonable to adopt a ``standard''
 kinetic energy of the order $E_{K}(10^{51}$ ergs$)=1$ for these
 two SNe. For SN 2006gi, some information might be inferred from the only two,
 until now published spectra. Indeed, the {\bf NOT} 
 spectrum\footnote{can be found 
 at http://www.supernovae.net/sn2006/sn2006gi.pdf} obtained  
  around maximum displays lower expansion velocities relative to
 SNe Ib 1990I and 1984L (as seen from the absorption troughs of He I and Fe II 
  optical lines). A similar conclusion, i.e low expansion velocity, is
 drawn from the nebular spectrum displayed by Taubenberger et al. 2009
 by means of the reported FWHM of the [O I] 6300,6364 \AA$~$ line compared 
 to a large sample of SESNe. Combining these facts with the noted fast decline 
 of the bolometric light curve, a late $e-$folding 
 decay time of 62 days, might lead one to speculate that the kinetic energy 
 is lower than a ``regular'' value of $E_{K}(10^{51}$ ergs$)=1$. 

%%%%%%%%%%%%%%%%%%%%%%%%%%%%%%%%%%%%%%%
\begin{figure*}
\centering
 \includegraphics*[width=15.5cm,height=14cm]{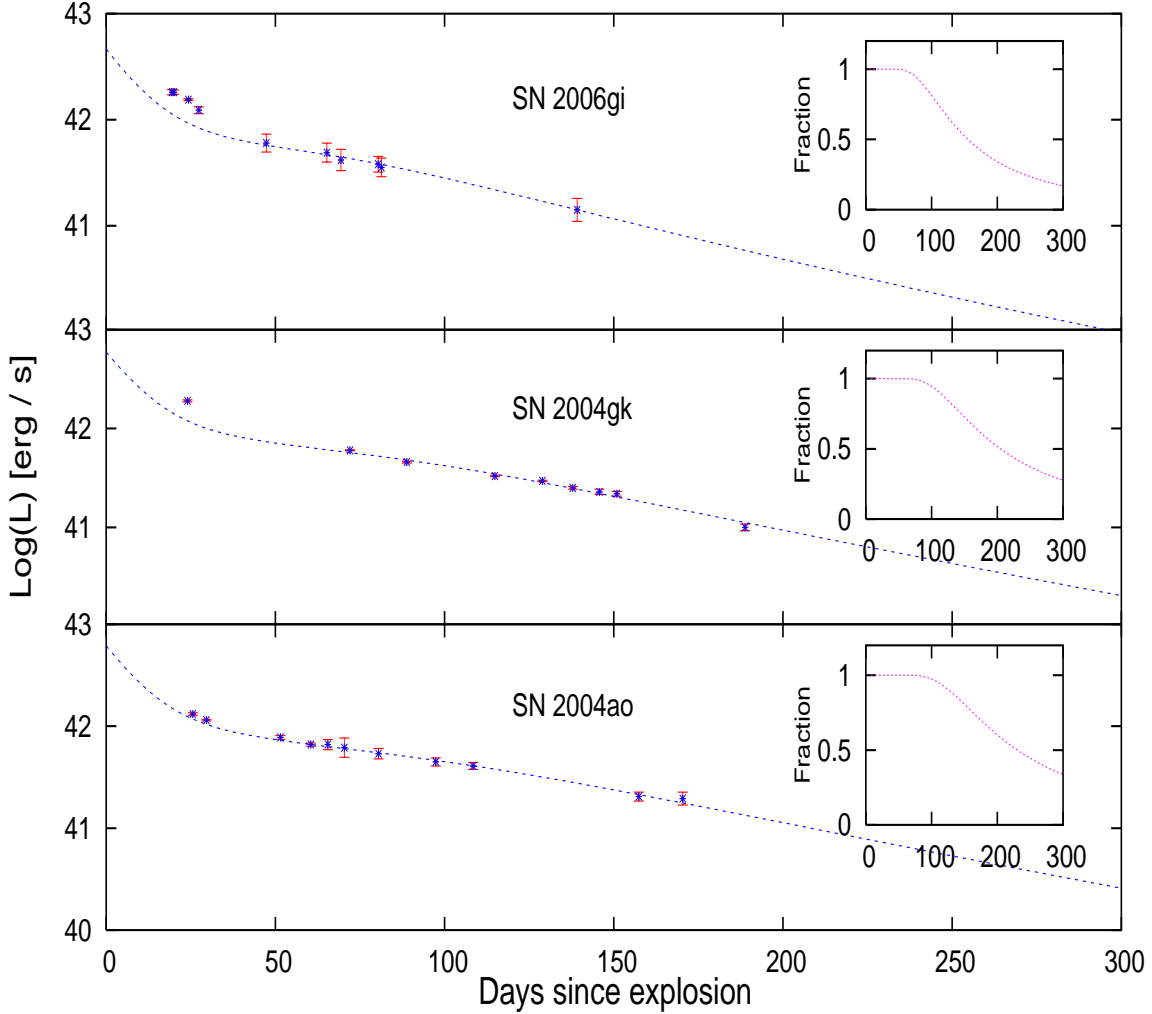}
 \figcaption{The computed $``BVRI''$ bolometric light curve of SN 2004ao(lower
 panel), SN 2004gk(middle panel) and SN 2006gi(upper panel). 
 An adopted $''JHK''$ contribution, of the order of 35$\%$, is added to the 
plotted luminosities. The best fits with the simplified $\gamma$-ray 
deposition model from $^{56}$Ni$\rightarrow$$^{56}$Co$\rightarrow$$^{56}$Fe 
are shown (see text). The windows display the evolution of the fraction of 
the $\gamma$-ray luminosity deposited in the envelope corresponding to the
best fits.} 
 \end{figure*}
%%%%%%%%%%%%%%%%%%%%%%%%%%%%%%%%%%%%%%%%%%%

  As mentioned above, restricting the energy to $E_{K}(10^{51} $ergs$)=1$ our 
 favored model fit indicates the following values:  SN 2004ao 
 $(M(^{56}Ni)=0.085 ~M_\odot$ 
 and $M_{ej}=6.3~ M_\odot )$; and SN 2004gk $(M(^{56}Ni)=0.082 ~M_\odot$ and 
 $M_{ej}=5.6~ M_\odot )$. For SN 2006gi, using  $E_{K}(10^{51} $ergs$)=0.5$ 
 gives $M(^{56}Ni)=0.064 ~M_\odot$ and $M_{ej}=3~ M_\odot$.
 In Fig. 11 we illustrate the best fit corresponding to these data.
  The windows in the panels of figure 10 highlight 
 the evolution of the fraction of the $\gamma-$ray luminosity deposited into 
 the envelope corresponding to our best fits.

  Adopting an energy of $E_{K}(10^{51}$ergs$)=1.5$ for SNe 2004ao and 2004gk
  and not changing $\gamma$-ray optical depth would increase the derived 
 ejecta masses by about 15-20$\%$.

  Though simple, the model 
 reproduces reasonable fits especially  the initial part of the radioactive 
  tail. In the [50-120] days time interval SN 2004ao and SN 2004gk are
  found to have almost identical luminosities (similar  $^{56}Ni$ masses), 
  while later on SN 2004ao seems to evolve higher and more slowly 
 (greater $M_{ej}$).
  The recovered SN 2006gi luminosities and the model fit require both the 
  lowest nickel and ejecta masses compared to the other two events.   

%%%%%%%%%%%%%%%%%%%%%%%%%%%%%%%%%%%%%%
\section{Summary and conclusions}
%%%%%%%%%%%%%%%%%%%%%%%%%%%%%%%%%%%%%%%
 
 We have presented $B,V,R,I$ optical photometry of three ESSNe, namely 
 SN Ib 2004ao, SN Ic 2004gk and SN Ib 2006gi covering about 200 
 days of evolution. Constraints on explosion dates, distance moduli and 
 total reddening are given and adopted throughout the paper.
 Colour evolution and light curve similarities and differences with
 some well-studied type Ib/c objects are highlighted and discussed.
 The observed $V$-light curve of SNe 2004ao displays a three-phase behaviour,
 seen also in SN 2004gk if considering the deviation of the last observational
 data around day 200. 

 The broad-band photometry is used to construct the quasi-bolometric light
 curves after adopting a constant contribution from $NIR$ part of the spectrum.
 The resulting light curves are compared to those of other hydrogen 
 deficient SNe, all constructed following the same method. 
 Around day 25 SNe Ib 2004ao and 2006gi are found to
 have similar luminosity, whereas thereafter at age 70 days 2006gi has
 $\sim$0.18 dex fainter luminosity compared to that of SN 2004ao.  
 As far as the bolometric light curve shape is concerned, SN 
 2006gi is most similar to SNe 1999dn and 1999ex, belonging to the fast 
 declining type Ib class. SN 2004ao appears to fit within the slow declining 
 Ib subgroup. SNe 2004ao and 2004gk are found to
 converge to similar luminosities in the time interval [50-100] days, while
 at early post-maximum phase SN 2004gk is brighter.  
 
 A simple $\gamma-$ray deposition
 model is described and then applied to estimate the main physical parameters
 of our sample. After adopting constraints on the explosion kinetic energy
 from the available published spectroscopy observations, 
 SN 2006gi appears to have both the lowest nickel and ejecta masses 
 compared to SNe 2004ao and 2004gk.
 
 A comparison of the resulting quasi-bolometric light curves for the three
  SNe, shown in Fig. 10 and Fig. 11, reveals the following. 
 1.All $e-$folding decay times are faster than the 
 $^{56}$Co decay time, i.e. 111.3 d,  in the time interval [50-150] days.
 2.The decay times appear to correlate with absolute luminosity, such that 
 the faintest SN 2006gi has a decay time of 62 days, while SN 2004ao the 
 has a decay time of 90 days and SN 2004gk has a decay time of 
 80 days. 3. SN 2004gk shows a third phase starting near or before 
 150 days.

  On the one hand, this observed late deficit in SN 2004gk is also reported 
 in both optical light curves and $B-V$ colour evolution, 
 reminiscent of what seen in type Ib SN 1990I (Elmhamdi et al. 2004). In
 SN 1990I however the sudden drop was also associated to a spectral blueshift
 (seen in the [O I] 6300,6364 \AA$~$ and [Ca II] 7291,7324 \AA $~$doublets),
 and explained in the framework of dust condensation as the ejecta expand
 and cool causing a drop in temperature below the threshold where dust can
 form (Elmhamdi et al. 2004) and absorb and block shorter wavelength radiation.
 
   On the other hand, from the spectroscopic point of view, nebular lines of 
  SN 2004gk, especially [O I] 6300,6364 \AA$~$ and 
  [Ca II] 7291,7324 \AA $~$doublets, show only a small shift to the blue
  ($\sim$10 \AA$~$ around day 200 corresponding to $\sim$500 km.s$^{-1}$ ;  
  Robert Quimby, Private communication).
  However in Modjaz et al. 2008, the published late time spectrum 
 of SN 2004gk, taken at 
  260 days, displays a single-peaked [O I] 6300,6364 \AA$~$ profile
  with no evidence for a blueshift deviation.
  Various mechanisms have been
  invoked to explain the line shape nature and 
  properties of the nebular [O I] 6300,6364 \AA$~$line in SESNe 
  ( i.e. moderate and extreme
   shifts in single- and double-peaked line profile cases; See Modjaz et 
   al. 2008 and Taubenberger et al. 2009 for detailed discussions). 
   Taubenberger et al. propose residual opacity effects as the most 
    likely reason for less extreme [O I] 6300,6364 \AA$~$ line blueshifts. 
  Moreover, dust should manifest its presence in a line blueshift 
  amount increasing in time, while that seems not be the case for SN 2004gk.
               
  It should be noted that so far only two
  type Ib/c events have showed observational evidence of possible 
  dust formation,
  namely SN Ib 1990I and SN Ib 2006jc. Based on evidences
  from line profiles and
  optical-infrared light curves (seen as early as day 50 since maximum)  
  the event was a clear case where dust formed or was present
  in the CSM and not the envelope. Indeed while the broad lines from the 
 envelope were not shifted, the narrow He I line from the CSM was   
 (Smith, Foley and 
  Filippenko 2008; Nozawa et al. 2008; Mattila et al. 2008)

 The principal goal of the present work has been to enrich the photometric
 observations of the hydrogen deficient class of core collapse SNe for future
 sample studies. 
 Indeed, such analyzes are stimulating for understanding common and 
 diverse physical aspects within the members of the sample. 
 Worth considering here is our discussion about the
 dust formation correlated effects in 2004gk (as being one of the rare
 Ibc SNe showing such possibility).
  As a consequence of the present paper we encourage future quantitative
 investigations of dust formation in SESNe. This
 would help us better understand this subclass and their environments
 and might and might help
 explain the origin of interstellar dust considering that core collapse SNe 
 are among the dust sources in the universe.

 Clearly, much is still to be done in order to eventually disclose new 
 frontiers in both observations and theory of SESNe.

{\bf Acknowledgments:} 

 The work of A. Elmhamdi was supported by KSU College of Science-Research
 center project No (Phys/2009/18). The work of D.Tsvetkov 
 was partly supported by 
 the Russian Leading Scientific Schools Foundation under grant NSh.433.2008.2 
 and by the RFBR grant 10-02-00249a. We would like to thank 
 R. Kirshner, R. Quimby, M. Modjaz, T. Matheson and C. Wheeler
 for helpful comments 
 and stimulating discussions about SNe 2004ao and 2004gk observations.
 A. Elmhamdi thanks V.Stanishev and T. Purismo (at NOT observatory) for
 the insightful discussions about SN 2006gi.
 The authors are grateful to S.Yu. Shugarov, who made some of the observations.
 A. Elmhamdi thanks S. Benetti for the useful data and comments on SN 1999dn.
 We acknowledge the usage of the HyperLeda and NED databases. 
 We thank the referee for the very helpful and constructive suggestions.

%%%%%%%%%%%%%%%%%%%%%%%%%%%%%%%%%%%%%
\end{document}